\documentclass[ALICE,manyauthors]{cernphprep}

\usepackage[utf8x]{inputenc}
\usepackage{graphicx}
\usepackage{lineno}
\usepackage[margin=1.0in]{geometry}
\usepackage{xcolor}
\usepackage{xspace}
\usepackage{multirow}

\usepackage[sort&compress,numbers]{natbib}
\usepackage{doi}
\usepackage{hyperref}

\graphicspath{{fig/}}

\newcommand {\pt}        {\ensuremath{p_{\mathrm{\textsc{t}}}}}

\newcommand {\meanptsq}  {\ensuremath{\langle p_{\mathrm{\textsc{t}}}^{2} \rangle}}
\newcommand {\pttrig}    {\ensuremath{p_{\mathrm{\textsc{t}}}^{\mathrm{trig}}}}
\newcommand {\y}         {\ensuremath{y}}
\newcommand {\sqrtSnn}   {\ensuremath{\sqrt{s_{_{\mathrm{NN}}}}}}
\newcommand {\sqrtSnnE}[2][TeV]  {$\sqrtSnn = #2\,\mathrm{#1}$}
\newcommand {\sqrtS}     {\ensuremath{\sqrt{s}}\,}
\newcommand {\sqrtSE}[2][TeV]  {$\sqrtS = #2\,\mathrm{#1}$}
\newcommand {\ee}        {\ensuremath{e^+e^-}}
\newcommand {\mumu}      {\ensuremath{\mu^+\mu^-}}
\newcommand {\mumuMB}    {\ensuremath{\mu \mu \mathrm{MB}}}
\newcommand {\Raa}       {\ensuremath{R_\mathrm{AA}}}
\newcommand {\Taa}       {\ensuremath{\langle T_\mathrm{AA} \rangle}}

\newcommand {\Npart}     {\ensuremath{\langle N_{\mathrm{part}} \rangle}}
\newcommand {\jpsi}      {\ensuremath{\mathrm{J}\kern-0.02em/\kern-0.05em\psi}}
\newcommand {\psip}      {\ensuremath{\psi\mathrm{(2S)}}}
\newcommand {\chic}      {\ensuremath{\chi_{c}}}
\newcommand {\pp}        {\ensuremath{\mathrm {p\kern-0.05em p}}}                                                                                
\newcommand {\PbPb}      {\ensuremath{\mathrm{Pb\mbox{-}Pb}}}
\newcommand {\AuAu}      {\ensuremath{\mathrm{Au\mbox{-}Au}}}
\newcommand {\nucnuc}    {\ensuremath{\mathrm{A\mbox{-}A}}}
\newcommand {\pPb}       {\ensuremath{\mathrm{p\mbox{-}Pb}}}
\newcommand {\dEdx}      {\ensuremath{\mathrm{d}E/\mathrm{d}x}}
\newcommand {\mee}       {\ensuremath{\mathrm{m}_{\ee}}}
\newcommand {\gev}       {\ensuremath{\,\mathrm{GeV}}}
\newcommand {\tev}       {\ensuremath{\,\mathrm{TeV}}}
\newcommand {\gevc}      {\ensuremath{\,\mathrm{GeV}\kern-0.05em/\kern-0.02em c}}
\newcommand {\gevcc}     {\ensuremath{\,\mathrm{GeV}\kern-0.05em/\kern-0.02em c^2}}
\newcommand {\cc}        {\ensuremath{{\mathrm c}\bar{{\mathrm c}}}}
\newcommand {\aeps}      {\ensuremath{A\kern-0.1em\times\kern-0.1em\varepsilon}}

\begin{document}

\begin{titlepage}

\PHnumber{2013-203}                 
\PHdate{October 30, 2013}              
%
%


\title{Centrality, rapidity and transverse momentum dependence of  \\
\jpsi\  suppression in \PbPb\ collisions at \sqrtSnnE{2.76}}
\ShortTitle{\jpsi\ suppression in  \PbPb\ collisions at \sqrtSnnE{2.76}}   
%
\Collaboration{ALICE Collaboration%
         \thanks{See Appendix~\ref{app:collab} for the list of collaboration members}}
\ShortAuthor{ALICE Collaboration}      

\begin{abstract}

The inclusive \jpsi\ nuclear modification  factor (\Raa) in \PbPb\ collisions at \sqrtSnnE{2.76} has been measured by  ALICE  as a function of centrality in the \ee\ decay channel at mid-rapidity ($|\y| < 0.8$) and as a function of centrality, transverse momentum  and rapidity in the  \mumu\ decay channel at forward-rapidity ($2.5 < \y < 4$).
The \jpsi\ yields measured in \PbPb\  are suppressed compared to those in \pp\ collisions scaled by the number of binary collisions. 
The \Raa\ integrated over a centrality range corresponding to 90\% of the inelastic \PbPb\ cross section is $0.72 \pm 0.06 (\mathrm{stat.}) \pm 0.10 (\mathrm{syst.})$ at mid-rapidity and  $0.58 \pm 0.01  (\mathrm{stat.}) \pm 0.09  (\mathrm{syst.})$ at forward-rapidity.
At low transverse momentum, significantly larger values of \Raa\ are measured at forward-rapidity compared to measurements at lower energy. 
These features suggest that a contribution to the \jpsi\ yield originates from charm quark (re)combination in the deconfined partonic medium.

\end{abstract}
\end{titlepage}

\setcounter{page}{2}

\section{Introduction}
\label{sec:intro}

The theory of Quantum Chromodynamics (QCD) predicts that the hot and dense nuclear matter produced  during the collision of ultra-relativistic heavy nuclei  behaves as a deconfined Plasma of Quarks and Gluons (QGP). 
This phase of matter exists for only a short time before the fireball cools down and the process of hadronization takes place. 
Heavy quarks are an important probe of the QGP since they are  expected to be produced only during the initial stage of the collision in hard partonic interactions,  thus experiencing the entire evolution of the system. 
It was predicted that in a hot and  dense deconfined medium like the QGP,  bound states of charm (c) and anti-charm ($\bar{\mathrm{c}}$) quarks, i.e.~charmonia,  are suppressed due to the screening effects  induced by the high density of color charges~\cite{Matsui:1986dk}. 
The relative production probabilities of charmonium states with different binding energies may provide important information on the properties of this medium and, in particular, on its temperature~\cite{Karsch:1990wi,1615967}.
Among the charmonium states,  the strongly bound \jpsi\ is of particular interest. 
The  \jpsi\ production  is a  combination from  prompt and non-prompt sources. 
The prompt \jpsi\ yield consists of the sum of direct \jpsi\ ($\approx$ 65\%) and  excited \cc\ states such as \chic\ and \psip\ decaying into \jpsi\ + X ($\approx$ 35\%)~\cite{Faccioli:2008ir}.
These excited  states have a smaller binding energy than the \jpsi. 
Non-prompt \jpsi\  production is directly related to beauty hadron production whose relative contribution increases with the energy of the collision. 
Experimentally, \jpsi\ production was studied in heavy-ion collisions at the Super Proton Synchrotron (SPS) and at the Relativistic Heavy Ion Collider (RHIC), covering a large energy range from about 20 to 200\gev\ center-of-mass energy per nucleon pair (\sqrtSnn).
A suppression of the inclusive \jpsi\ yield  in  nucleus--nucleus (\nucnuc) collisions with respect to the one measured in proton-proton (\pp) scaled by the number of binary nucleon-nucleon collisions was observed. 
In the  most central events, the  suppression is beyond the one induced by cold nuclear matter effects (CNM), such as shadowing and nuclear absorption, at both SPS~\cite{Alessandro:2004ap,Arnaldi:2009ph} and RHIC~\cite{Adare:2007gn}.
At the SPS the \jpsi\ suppression is compatible with the melting of the excited states whereas the RHIC data suggest a small amount of suppression for the direct \jpsi\ ~\cite{Adare:2006ns,Adare:2011yf}.
Similar predictions on sequential suppression~\cite{1615967} were made for the bottomonium family, which has become accessible at the Large Hadron Collider (LHC)  energies. 
The sequential suppression of  the $\varUpsilon$(1S), $\varUpsilon$(2S) and $\varUpsilon$(3S) states was first observed by the CMS experiment in \PbPb\ collisions at \sqrtSnnE{2.76}~\cite{Chatrchyan:2012lxa}.

The first ALICE measurement of the inclusive \jpsi\ production in central \PbPb\ collisions at \sqrtSnnE{2.76} at forward-rapidity has  shown less suppression
compared to PHENIX results in central \AuAu\ collisions at \sqrtSnnE{0.2}~\cite{Abelev:2012rv}. 
At \sqrtSnnE{2.76}, the charm quark density produced in the collisions increases  with respect to SPS and RHIC energies~\cite{Abelev:2012vra}.
This may result in the enhancement of the probability to create \jpsi\ mesons from  (re)combination of charm 
quarks~\cite{{BraunMunzinger:2000px},{Thews:2000rj}}.
If the \jpsi\ mesons are fully suppressed in the QGP, their creation will take place at chemical freeze-out (near the phase boundary)
as detailed in \cite{{BraunMunzinger:2000px},{Gorenstein:2000ck},{Andronic:2007bi}}.
If \jpsi\ mesons survive in the QGP, production may take place continuously during the  QGP lifetime 
\cite{{Thews:2000rj},{Zhao:2007hh},{Liu:2009nb}}. 
Because of the large increase of the \cc\ cross-section towards LHC energy the (re)combination mechanism may become dominant there.
According to statistical~\cite{{BraunMunzinger:2000px}} and partonic transport~\cite{{Zhao:2007hh},{Liu:2009nb}} models,  this contribution leads to an increase of the \Raa\  at the LHC with respect to the one observed at RHIC.
In particular,  this scenario predicts an increase of  the \Raa\  from forward- to mid-rapidity, where the density of charm quarks is higher. 
Furthermore, in order to (re)combine, two charm quarks need to be close enough in phase space, so that low transverse momentum \jpsi\ production  is expected to be favored. 
The transverse momentum and rapidity dependence of the \jpsi\ \Raa\ are therefore crucial observables to sharpen the interpretation of the results, providing a deeper insight on the balance between \jpsi\ (re)combination and suppression. 

In this Letter, we present results on the nuclear modification factor for inclusive \jpsi\ in \PbPb\ collisions at \sqrtSnnE{2.76} as a function of collision centrality, transverse momentum and rapidity. 
Complementary to our results, \jpsi\ suppression at large transverse momentum in  \PbPb\ collisions, was reported previously by ATLAS~\cite{Aad:2010aa} and CMS~\cite{Chatrchyan:2012np}.

\section{Experimental apparatus and data sample}
\label{sec:experiment}

ALICE is a general purpose heavy-ion experiment.
A detailed description of the experimental apparatus can be found in~\cite{Aamodt:2008zz}. 
It consists of a central barrel covering the pseudo-rapidity interval $ |\eta| < 0.9 $ and a muon spectrometer  covering   $ - 4  <  \eta < - 2.5 $~\footnote{In the ALICE reference frame,  the muon spectrometer covers a negative $\eta$ range and consequently a negative $y$ range. We have chosen to present our results with a positive \y\ notation.}.  
\jpsi\ production is measured in both rapidity ranges: at mid-rapidity 
in the dielectron decay channel and at forward-rapidity in the dimuon decay channel. 
In both cases the \jpsi\ transverse momentum (\pt) coverage extends down to zero.

At mid-rapidity, the detectors used for the \jpsi\ analysis are the Inner Tracking System (ITS)~\cite{2010JInst...5.3003A} and the Time Projection Chamber (TPC)~\cite{2010NIMPA.622..316A}.
The ITS is composed of six concentric cylindrical layers of silicon detectors with radii ranging from 3.9 to 43$\,$cm with respect to the beam axis.
Its main purpose is to provide the reconstruction of the primary interaction vertex as well as secondary decay vertices of heavy flavored particles.
In addition, the two innermost layers can provide an input at level zero (L0) to the trigger system.
The TPC, with an active volume extending from 85 to 247$\,$cm in the radial direction, is the main tracking detector of the central barrel and also provides particle identification via the measurement of the specific energy loss (\dEdx) in the detector gas.

At forward-rapidity,  the \jpsi\ analysis is carried out using  the muon spectrometer~\cite{Aamodt:2011gj}. 
The spectrometer consists of a ten interaction length front absorber, filtering the  muons in front of five  tracking stations  made of two planes of cathode pad chambers each. 
The third station is located  inside a dipole magnet with a 3$\,$Tm field integral. 
The spectrometer is completed by a Muon Trigger system (MTR)  made of two stations,  each equipped with two planes of resistive plate chambers. 
The trigger chambers are placed behind a 1.2$\,$m thick iron wall to stop secondary hadrons escaping from the front absorber and low momentum  muons coming mainly from $\pi$ and K decays.
Throughout its full length, a conical absorber made of tungsten, lead and steel protects the muon spectrometer against secondary particles generated by the interaction with the beam pipe
of primary particles produced at large $\eta$.

Additional forward detectors, the VZERO~\cite{Abbas:2013taa} and the Zero Degree Calorimeters (ZDC)~\cite{ALICE:2012aa}, are used for triggering and event characterization.
The VZERO detector is composed of two scintillator arrays, 32 channels each, placed on both sides of the Interaction Point (IP). 
It covers $2.8 \leq \eta \leq 5.1$ (VZERO-A) and $- 3.7 \leq \eta \leq - 1.7$ (VZERO-C).
The ZDC are located at  a distance of 114$\,$m  on both sides of the IP and can detect spectator neutrons and protons.

The results presented in this Letter are based on data collected during the 2010 
and 2011 LHC \PbPb\ runs for the dielectron analysis and on data collected in the 2011 run for 
the dimuon one.
Forward-rapidity results in the dimuon channel from the 2010 data set, based on an integrated luminosity about 25 times  smaller than the 2011 data set, have been published previously in~\cite{Abelev:2012rv}.
The minimum bias (MB) trigger for the 2011 data set is defined by the coincidence of signals in the two VZERO arrays synchronized with the passage of two crossing Pb bunches. 
In the 2010 data set, the MB trigger had an additional requirement on hits in the ITS. 
The two MB trigger definitions, however, lead to very similar trigger efficiencies, which are larger than 95\% for inelastic \PbPb\ collisions. 
Electromagnetic interactions are rejected at the level one trigger (L1) by applying a cut on the minimum energy deposited by spectator neutrons in the ZDC.
Beam induced background is further reduced at the offline level by applying timing cuts on the signals from the VZERO and ZDC detectors.

At mid-rapidity, the 2010 data sample used in the electron analysis consists of 15 million events collected with the MB trigger, corresponding to  an integrated luminosity of $2.1 \, \mu \mathrm{b}^{-1}$. 
The 2011 event sample was enriched with central and semi-central \PbPb\ collisions by using thresholds on the VZERO multiplicity at the L0 trigger.
The inspected integrated luminosity amounts to $25.6 \, \mu \mathrm{b}^{-1}$, out of which we analyzed 20 million central (0\%--10\% of the centrality distribution) and 20 million semi-central (10\%--50\%) events. 
The summed 2010 and 2011 datasets correspond to an integrated luminosity of $\mathcal{L}_{\rm int} = 27.7 \pm 0.4 (\mathrm{stat.}) \, ^{+2.2}_{-1.8} (\mathrm{syst.} \; \sigma_{\PbPb}) \, \mu \mathrm{b}^{-1}$.
At forward-rapidity, the 2011 data sample is made of about 17 million \mumuMB\ triggers. The \mumuMB\ trigger is defined as the occurrence of the MB condition in coincidence with the detection in the MTR of two opposite-sign muons tracks. 
The MTR is capable of (i) delivering L0 trigger decisions at 40$\,$MHz based on the detection of one or two muon trigger tracks, (ii) computing an approximate value of the transverse momentum of muon trigger tracks (\pttrig) and (iii) applying a threshold\footnote{The threshold is defined as \pttrig\ for which the trigger probability is 50\% and does not lead to a sharp cut in \pt.} on the \pttrig. A 1\gevc\ threshold, applied on both muons, was chosen to collect this data sample. 
A scaling factor $F_{\mathrm{norm}}$  is used to obtain the number of equivalent MB events from the number of \mumuMB\ ones.
It is defined as the ratio, in a MB data sample, of the number of MB events divided by the number of events fulfilling the \mumuMB\ trigger condition.
Its value, averaged over the entire data sample, is 
$F_{\mathrm{norm}} = 30.56 \pm 0.01 (\mathrm{stat.}) \pm 1.10 (\mathrm{syst.})$.
The integrated luminosity used in this analysis is therefore $\mathcal{L}_{\rm int} =   N_{\mumuMB} \times  F_{\mathrm{norm}} / \sigma_{\PbPb}  = 68.8 \pm 0.9 (\mathrm{stat.})  \, \pm 2.5 (\mathrm{syst.} \; F_{\mathrm{norm}}) \,  ^{+5.5}_{-4.5} (\mathrm{syst.} \; \sigma_{\PbPb})  \, \mu\mathrm{b}^{-1}$ 
assuming  an inelastic \PbPb\ cross-section $\sigma_{\PbPb} = 7.7 \pm 0.1 \, ^{+0.6}_{-0.5}\, \mathrm{b}$~\cite{ALICE:2012aa}. 

The centrality determination is based on a fit to the VZERO amplitude distribution as described  in~\cite{Abelev:2013qoq}.
The fit, based on the Glauber model, allows for the extraction of  collision-related variables such as the average of  number of participant nucleons  \Npart\ and the average of the nuclear overlap function \Taa\ per centrality class. 
Numerical values are given in Table~\ref{tab:taa}.
\begin{table}
\caption{\label{tab:taa} The average of  number of participating nucleons \Npart\  and the  average value of the nuclear overlap function \Taa\ with their associated systematic uncertainty 
for the  centrality classes, expressed in percentages of the nuclear cross-section~\cite{Abelev:2013qoq}, used in these analyses.}
\begin{center}
\begin{tabular}{c|ll}
\hline
\hline
Centrality    & \Npart\              &  \Taa\ (mb$^{-1}$) \\
\hline
0\%--10\%     & $356.0 \pm 3.6 $           &  $23.44 \pm 0.76  $    \\
10\%--20\%    & $260.1 \pm 3.8 $           &  $14.39 \pm 0.45 $     \\
20\%--30\%    & $185.8 \pm 3.3$            &  $8.70  \pm 0.27 $     \\
30\%--40\%    & $128.5 \pm 2.9$            &  $5.00  \pm 0.18 $     \\
40\%--50\%    & $ 84.7 \pm 2.4$            &  $2.68  \pm 0.12 $     \\
50\%--60\%    & $ 52.4 \pm 1.6$            &  $1.317 \pm 0.071$     \\
60\%--70\%    & $ 29.77 \pm 0.98$          &  $0.591 \pm 0.036$    \\
70\%--80\%    & $ 15.27 \pm 0.55$          &  $0.243 \pm 0.016$    \\
80\%--90\%    & $  7.49 \pm 0.22$          &  $0.0983\pm 0.0076$    \\
\hline
0\%--20\%     & $308.1 \pm 3.7$            &  $18.91 \pm 0.61 $     \\
10\%--40\%    & $191.5 \pm 3.3 $           &  $9.36  \pm 0.30$      \\
40\%--90\%    & $37.9 \pm 1.2$             &  $0.985 \pm 0.051 $   \\
0\%--90\%     & $124.4 \pm 2.2$            &  $6.27  \pm 0.21$       \\
\hline
\hline
\end{tabular}
\end{center}
\end{table}                        
Both the electron and muon analyses were carried out on an event sample corresponding to the most central  90\% of the inelastic \PbPb\ cross-section.  In this centrality range the efficiency of the MB trigger is 100\% and the contamination from electromagnetic processes is negligible.

\section{Data analysis}
\label{sec:data_analysis}

\jpsi\ candidates are formed by combining pairs of  opposite-sign (OS) electron and muon tracks reconstructed in the central barrel and in the muon spectrometer, respectively.

Electron candidates are selected by cutting  on the quality of tracks reconstructed  in the ITS and the TPC.
The selection criteria are very similar to those used in the previous analysis of \pp\  collisions at \sqrtSE{7}~\cite{Aamodt:2011gj} using  a tighter selection on electron identification.
A hit in one of the two innermost layers of the ITS is required.
This rejects a large fraction of background resulting from photon conversions  in the detector material.
The tracks are required to have at least 70 out of a maximum of 159 clusters in the TPC 
and to pass a quality cut based on the  $\chi^2$ of the TPC track fit divided by the number of clusters attached to the track. 
Electron identification is done using the TPC, requiring the \dEdx\ signal to be compatible with the electron expectation  within a band of $(-2.0;+3.0) \sigma$ or $(-1.5;+3.0) \sigma$ for the 2010 or 2011 data, respectively, where $\sigma$ denotes the resolution of the \dEdx\  measurement. 
Due to a lower \dEdx\ resolution for the 2011 data, for $|\eta|<0.5$ a more restrictive electron selection, $(-0.9;+3.0)\sigma$, is applied.
The electron/hadron separation is further improved by rejecting tracks which are compatible with the pion expectation within $3.5 \sigma$ and with the proton expectation within $3.5 \sigma$ or $4.0 \sigma$ in 2010 or 2011 data, respectively.
Since the electrons from a \jpsi\ decay have a momentum of 1.5\gevc\ in the mother particle rest frame,  a cut of \pt\ $> 0.85$\gevc\  on the candidate tracks is applied to reject the combinatorial background from low momentum electrons. 
Finally, to ensure good tracking and particle identification in the TPC, only candidates within $|\eta| <0.8$ are selected. 

Muon tracks are reconstructed in the  muon spectrometer as detailed in~\cite{Aamodt:2011gj} for pp collisions. This  procedure remains basically unchanged for \PbPb\ collisions. 
However, to cope with the  large background in central events, some selection criteria were  tightened compared to the \pp\ analysis. 
The search area for finding clusters  associated to tracks is reduced by a factor of nine, both muon candidates have to match a track segment in the trigger chambers and the track pseudo-rapidity has to be in the range $-4<\eta<-2.5$.
A further cut on the track transverse coordinate at the end of the front  absorber ($R_{\mathrm{abs}}$) is applied ($17.6 \leq  R_{\mathrm{abs}} \leq 89.5 \, \mathrm{cm}$) to ensure that  muons emitted at small angles, i.e. those that have crossed a significant fraction of the thick  beam shield, are rejected. 
Finally, to remove events very close to the edge of the spectrometer acceptance, only muon pairs in the rapidity range $2.5 < \y < 4$ are accepted.

\begin{figure}[h!]
\begin{center}
\includegraphics[width=1\textwidth]{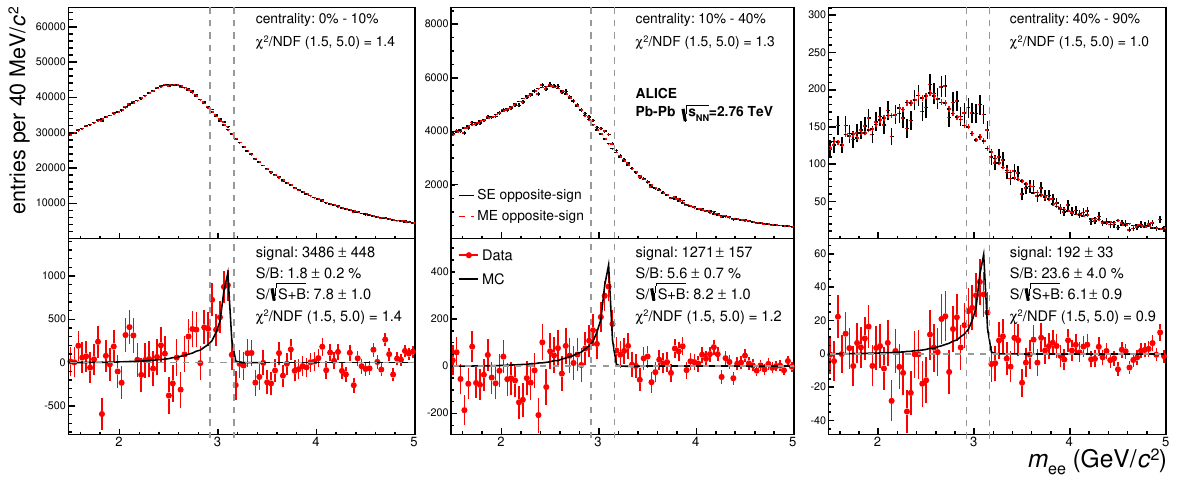} 
\end{center}
\caption{Top panels: invariant mass distributions for opposite-sign (OS) and mixed-events (ME) electron pairs. 
Bottom panels: OS invariant mass spectra, after the subtraction of the ME distributions, 
with a comparison to the Monte Carlo signal (solid lines) superimposed. The MC signal is scaled to match the
integral of the OS distribution within the mass counting window.
From left to right, distributions correspond to the centrality ranges 0\%--10\%,10\%--40\% and 40\%--90\%, respectively.
The panels for the 0\%--10\% and 10\%--40\% centrality ranges are obtained from the 2011 data, while the ones for the 40\%--90\%
centrality range are obtained from the 2010 data.
The two vertical dashed lines shown in each panel indicate the mass interval used for signal counting.}
\label{fig:invmassee}
\end{figure}

In the \ee\ decay channel, the \jpsi\ yields are extracted by counting the number of entries  in the invariant mass range $2.92 <  \mee < 3.16 $\gevcc\  after subtracting the  combinatorial background.
Due to the radiative decay channel and the energy loss of the electrons in the detector material via bremsstrahlung, only $\approx$68\% of the \jpsi\ are reconstructed with the mass in the counting mass interval. 
The background shape is obtained using the mixed-event (ME) technique.
Uncorrelated lepton pairs are created from different \PbPb\ events that have similar global properties such as centrality, primary vertex position and event plane angle.  
The background shape from ME is scaled to match the same-event (SE)  invariant mass distribution  in the ranges $1.5< \mee  <2.5$\gevcc\ and $3.2< \mee  <4.2$\gevcc.
These mass ranges were chosen such that they are close to the signal region 
and have equal number of bins on each side  of the signal region. 
The lower (1.5\gevcc) and upper (2.5\gevcc) limits of the first mass range are chosen in order to avoid sensitivity to correlated low-mass dielectron pairs and \jpsi\ bremsstrahlung tail, respectively.
The second mass range is limited by the upper limit of the signal region (3.2\gevcc)
and extends to 4.2\gevcc\ to match in size the first mass interval. 
Here the influence of the  \psip\ on the background matching procedure is neglected since the \psip\ dilepton yields are expected  to be roughly 60 times  smaller than \jpsi\ yields (estimation based on LHCb measurements of \jpsi\ \cite{Aaij:2011jh} and \psip\ \cite{Aaij:2012ag} cross-sections in \pp\ collisions at 7\tev).
A good matching between the SE and ME distributions is observed over a broad  mass range outside the \jpsi\ mass  region, as visible in the top panels of Fig.\,\ref{fig:invmassee}. 
This is a clear sign that the contribution of correlated pairs to the OS mass spectrum is small with respect to the uncorrelated background or has a similar shape.
The bottom panels of Fig.\,\ref{fig:invmassee} show the background-subtracted  invariant mass spectra compared to  the \jpsi\ signal shape from a Monte-Carlo (MC) simulation.
The bremsstrahlung tail from the electron energy loss in the detector material and the \jpsi\ radiative decay  channel (\jpsi\ $\to$ \ee\ $\gamma$) is well described in the MC.
As shown in Fig.\,\ref{fig:invmassee}, it is possible to study the \jpsi\ production in three centrality intervals (0\%--10\%,10\%--40\%,40\%--90\%), with a signal-to-background ratio (S/B), evaluated in the range $2.92 < \mee < 3.16$\gevcc , increasing from 0.02 to 0.25  from central to peripheral collisions.

\begin{figure}[h!]
\begin{center}
	\includegraphics[width=0.90\linewidth,keepaspectratio]{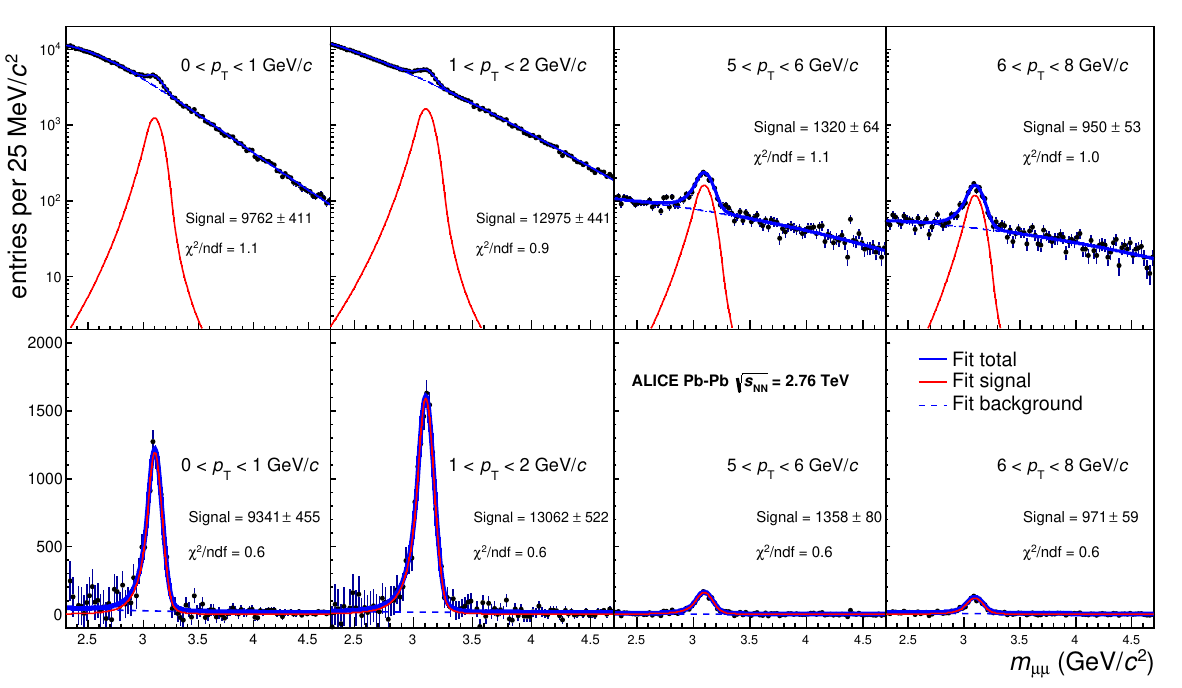}  
\caption{Top panels: fits of the dimuon invariant mass spectra in selected \pt\ intervals. Bottom panels: idem after subtraction of the combinatorial background with the event mixing technique. 
Distributions correspond to the centrality class 0\%--90\% and  $2.5< \y <4$.}
\label{fig:InvMassPerfVsPt}
\end{center}
\end{figure}

In the \mumu\ decay channel, the \jpsi\ raw yield is extracted in each centrality and kinematic interval by using two different methods.
In the first approach, the OS dimuon invariant mass distribution is fitted with the sum of an extended Crystal Ball (CB2) function to describe the signal, and a Variable Width Gaussian (VWG) function for the background.
The CB2 function extends the standard Crystal Ball (Gaussian plus power-law tail at low masses~\cite{CBdef}) by an additional power-law tail at high  masses with parameters independent of the low mass ones.
The VWG function is a Gaussian function with a fourth parameter to allow linear variation of the width with the invariant mass of the dimuon pair.
The \jpsi\ signal is clearly visible in all centrality, \pt\ or \y\ intervals even before any background subtraction, as can be observed in the top panels of Fig.~\ref{fig:InvMassPerfVsPt}, where examples of invariant mass spectra  fits in selected \pt\ intervals are shown.
The signal-to-background ratio, evaluated within 3 standard deviations with respect to the \jpsi\ pole mass, varies from 0.16 at low \pt\ up to 1.2 at high \pt . 
The corresponding values in centrality and \y\ intervals are in the range 0.16--6.5  from central to peripheral collisions and 0.19--0.59 from low to high rapidity.   
In all cases, the significance is larger than 10.
In the second approach, the combinatorial background was subtracted using an event-mixing technique. 
The background  shape obtained from ME was normalized to the data through a combination of the
measured like-sign muon pairs from SE.
Fig.~\ref{fig:InvMassPerfVsPt} (bottom  panels) shows the resulting mass distribution fitted with the sum of a CB2  and an exponential which accounts for residual correlated background.  
In both approaches, the position of the peak of the CB2 function ($m_{\jpsi}$), as well as its width ($\sigma_{\jpsi}$), are free parameters of the fit. 
Their values, obtained by fitting the invariant mass spectrum integrated  over \pt, \y\ and centrality, are  $m_{\jpsi}$ = $3.103 \pm 0.001\gevcc$ (shifted up by 0.2\% with respect to the PDG mass~\cite{Beringer:1900zz}) and $\sigma_{\jpsi}$ = $0.071 \pm 0.001\gevcc$. 
More details about the fitting procedures and the different parameters of the signal and background  line shapes are discussed in section \ref{sec:technics}.

The measured number of \jpsi\  ($N_{\jpsi}^{i}$) in a centrality class $i$ is normalized to the corresponding number of MB events falling in the centrality class ($N_{\mathrm{events}}^{i}$) and further corrected for the  branching ratio (BR) of the dilepton decay channel, the acceptance $A$ and the efficiency $\epsilon^{i}$  of the detector.
In the \mumu\ analysis, $N_{\mathrm{events}}^{i}$ is computed by multiplying the number of \mumuMB\ triggered events  by the $F_{\mathrm{norm}}$ factor (described in section \ref{sec:experiment}) scaled by the width of the centrality class $i$.
The  inclusive \jpsi\ yield  for the measured \pt\ and \y\ ranges is then given by:
\begin{eqnarray}
Y_{\jpsi}^{i}  =  \frac{N_{\jpsi}^{i}}{{\rm BR}_{\jpsi \rightarrow l^{+}l^{-}} N_{\rm{events}}^{i} \, A \times \epsilon^{i}  }.
\label{eq:jpsiyield}
\end{eqnarray}

The acceptance times efficiency product (\aeps) is defined as the ratio between the number of reconstructed \jpsi\ divided by the number of generated ones in the the kinematic range under study.   
In the \ee\ decay channel, \aeps\ is calculated from MC simulations. 
These MC events are a superposition of \PbPb\ collisions generated with an appropriate  HIJING~\cite{HIJINGref}  tune reproducing the measured charged particle density~\cite{Aamodt:2010cz} and \jpsi\ generated from  parametrized \pt\ and \y\ distributions (details in section~\ref{sec:technics}). 
The \jpsi\ dielectron decays are performed using PHOTOS \cite{PHOTOS:1991,PHOTOS:1994}. 
The particles are then transported through a simulation of the ALICE detector using GEANT 3.21 \cite{GEANTref}.
The geometrical acceptance is about 34\%.
The estimated integrated \aeps\ for the \jpsi\ emitted in $|y|<0.8$ amount to 0.080, 0.085 and  0.093 for the 0\%--10\%, 10\%--40\% and 40\%--90\% centrality classes in the 2010 data sample and 0.026 and 0.028 for the 0\%--10\% and 10\%--40\% centrality classes in the 2011 one, respectively. 
The large difference in efficiency is mainly due to  a lower efficiency of  the two innermost ITS layers  and  stronger particle identification cuts  used for the 2011 data set.
In the \mumu\ decay channel, \aeps\ was computed using an embedding technique where MC \jpsi\ particles are injected into the raw data of real events and then reconstructed. 
The MC \jpsi\ \pt\ and \y\ parametrization is taken from actual measurements.
PYTHIA~\cite{Sjostrand:2006za} takes care of the \jpsi\ decays and the daughter particles signals in the detector, given  by GEANT 3.21,  are then added to the real \PbPb\ events. 
When studied as a function of centrality, \aeps\ decreases by 7.9\%, from  0.127 in the 80\%--90\% centrality class to  0.117  in the  0\%--10\% one.
The centrality integrated  \aeps\ (0\%--90\% centrality class) is 0.120 with a negligible statistical uncertainty. 
The geometrical acceptance is about 18\%.
The quantity \aeps\ shows a non-monotonic dependence on \pt, starting at approximately 0.124 at zero transverse momentum, reaching a minimum of 0.103 at 1.5\gevc\ and then linearly increasing up to 0.264 at 8\gevc. 
The rapidity dependence of  \aeps\ reflects the geometrical acceptance  of the muon pairs with a maximum of 0.189 centered at $\y = 3.3$ decreasing towards the edges of the acceptance to 0.033 (0.060) at $\y = 2.5$ ($\y = 4.0$).

Finally, the nuclear modification factor is calculated as the ratio between the corrected \jpsi\ yield in \PbPb\ collisions 
$Y_{\jpsi}^{\PbPb}$ and the \jpsi\ cross-section in \pp\ collisions scaled by the nuclear overlap function:
\begin{equation}
\Raa = \frac{Y_{\jpsi}^{\PbPb}}{\Taa  \times \sigma_{\jpsi}^{\pp}}.
\end{equation}

\begin{table}[htp]
\caption{Inclusive \jpsi\ production cross-sections at mid-rapidity used in the interpolation procedure.
The \jpsi\ are assumed unpolarized and the systematic uncertainties do not include
the contribution from unknown polarization.
}
\centering
\begin{tabular}{c c c c c c c}
\hline
\hline
Experiment & Collision energy \sqrtS & Rapidity range & $B_{ll} d\sigma/d\y$ at $\y=0$ & stat. & syst. & Reference \\
           & (TeV)                            &                & (nb)                         & (nb)  & (nb)              &           \\
\hline 
PHENIX     & 0.2                              & $|\y|<0.35$     & 44.30                        & 1.40  & 6.80              & \cite{Adare:2006kf} \\
CDF        & 1.96                             & $|\y|<0.6$      & 201.6                        & 1.0   & $^{17.8}_{-16.3}$ & \cite{Acosta:2005}  \\
ALICE      & 2.76                             & $|\y|<0.9$      & 255.8                        & 58.7  & 45.9              & \cite{Abelev:2012kr} \\
ALICE      & 7                                & $|\y|<0.9$      & 409.9                        & 36.8  & 59.4              & \cite{Aamodt:2011gj,Aamodt:2011gjerr}\\
Interpolation & 2.76                          & $|\y|<0.8$      & 252.6                        & 16.4  & 25.8              & this work \\
\hline
\hline
\end{tabular}
\label{tab:interpolation}
\end{table}

In the dielectron analysis, the \pp\ reference was obtained by interpolating the inclusive \jpsi\ cross-sections at mid-rapidity measured by PHENIX~\cite{Adare:2006ns} at  \sqrtSE{0.2}, CDF~\cite{Acosta:2005} at \sqrtSE{1.96} and ALICE~\cite{{Abelev:2012kr},{Aamodt:2011gj},{Aamodt:2011gjerr}} at \sqrtSE{2.76 \; \mathrm{and} \; 7}. 
All the data points used in this procedure are listed in Table \ref{tab:interpolation}. 
The interpolation was done by fitting the data points with several functions assuming a linear, an exponential, a power law or a polynomial \sqrtS\ dependence. 
The value of the interpolated pp reference at mid-rapidity, $\mathrm{d}\sigma_{\jpsi}^{\pp}/\mathrm{d}\y = 4.25 \pm 0.28 (\mathrm{stat.}) \pm 0.43 (\mathrm{syst.}) \, \mu\mathrm{b}$,  is consistent with the one measured by ALICE \cite{Abelev:2012kr}, but the total uncertainty is twice smaller,
being driven mainly by the CDF result.
The statistical uncertainty was obtained from the fitting procedure, while the systematic one was obtained by changing the fit function and by shifting the data points within their experimental systematic uncertainties.
For the dimuon analysis, the \pp\ reference and its associated uncertainties are extracted from the ALICE measurement \cite{Abelev:2012kr}.

\section{Systematic uncertainties}
\label{sec:technics}

The main sources of systematic uncertainties for the \jpsi\ \Raa\ evaluation are the tracking efficiency, the  signal extraction procedure, the parameterization of the \jpsi\ kinematic distributions used as input for the MC simulations, the uncertainty on the nuclear
overlap function and the uncertainty on the \jpsi\ \pp\ cross-section at \sqrtSE{2.76}. 
Other analysis-dependent sources are detailed in the following. 
The systematic uncertainties have been evaluated as a function of centrality and,  for the dimuon analysis, of  \pt\ and \y.
An overview of systematic uncertainties is given in Table~\ref{tab:syst}.

In the dielectron analysis, the systematic uncertainties on the signal extraction are estimated by varying the mass region used to count the signal, the mass interval used for matching the ME background  and the OS distribution.
In all these cases, the background-subtracted OS distribution is compared to the MC signal shape and the normalized $\chi^2$ obtained is always close to 1. 
This shows that, after background subtraction, the number of correlated pairs not related to \jpsi\ decays  is small and does not induce a sizeable systematic uncertainty.
The centrality dependent systematic uncertainty on the signal extraction, taken as the RMS of the distribution of the number of \jpsi\ obtained from  all the performed tests, ranges from 7\% to 9\% and  4\% to 6\% for 2010 and 2011 data, respectively.
The systematic uncertainties due to track reconstruction and particle identification are evaluated by varying all the analysis cuts.
For each cut variation, the number of \jpsi\ signal counts is corrected with the corresponding \aeps . 
The RMS of this quantity is  found to vary in the range 6--9\% and 4--5\% in the 2010 and 2011 data, respectively. 
Since the signal extraction procedure must be used for every cut variation, the systematic uncertainties  due to analysis cuts and signal extraction cannot be truly disentangled.
Thus, a global systematic uncertainty is introduced as the RMS of the  distribution of corrected results when varying both signal extraction parameters and cut values. 
These systematic uncertainties range between 8\% and 11\% depending on the centrality interval. 
The central value for the corrected \jpsi\  yield is chosen to be the mean of the  distribution obtained from all the performed tests.

\begin{table}[htp]
\caption{Systematic uncertainties entering the \Raa\ calculation. The type I (II) stands   for correlated (uncorrelated) uncertainties within a given set of data points. The uncorrelated  systematic uncertainties  (type II) are given as a range.
}
\centering
\begin{tabular}{c|c c c c|c c}
\hline
\hline
Channel & \multicolumn{4}{c}{\mumu} &  \multicolumn{2}{|c}{\ee} \\
\hline
 & \multicolumn{2}{c}{Centrality} & \multicolumn{2}{c}{\pt\ or \y } & \multicolumn{2}{|c}{Centrality} \\
 & value (\%) & type & value (\%) & type &  value (\%) & type\\
\hline 
signal extraction         & 1--3         & II       & 1--5         & II       & \multirow{2}{*} {8-11}   &  \multirow{2}{*}{II} \\
tracking efficiency       & 11 and 0--1  & I and II & 1 and 8--14   & I and II &   &   \\
trigger efficiency        & 2 and 0--1   & I and II & 1 and 2--4    & I and II & n/a    &    \\
input MC parameterization & 3            & I        & 0--8          & II       & 5      &  I  \\
matching efficiency       & 1            & I        & 1             & II       & n/a    &    \\
centrality limits		  & 0--5         & II       & 0--1          & I        & 0--3   & II \\ 
\Taa\                     & 3--8         & II       & 3             & I        & 3--5   & II \\
$\sigma_{\jpsi}^{\pp}$    & 9            & I        & 6 and 5--6    & I and II & 12     & I  \\
$F_\mathrm{{norm}}$       & 4 		     & I		& 4 	  	    & I        & n/a    &    \\
\hline
\hline
\end{tabular}
\label{tab:syst}
\end{table}

In the dimuon analysis, the systematic uncertainty on the signal extraction is estimated by fitting the invariant mass distribution with and without background subtraction and by varying the  parameters that define the power law shapes at low and high masses of the CB2 signal function. 
These fit parameters  are not constrained by the data and cannot be let free during the fitting procedure. 
They have been fixed to different values extracted either from simulations or from \pp\ data, where the signal-to-background ratio is more favorable.
Fits corresponding to the various choices for the CB2 tails are performed, keeping the background parameters free, and varying the invariant mass range used for the fits. 
The raw \jpsi\ yield is determined as the average of the results obtained with the above procedure and the corresponding systematic uncertainty is defined as the RMS of the deviations from the average.	
As a function of centrality (\pt\ or \y) the  systematic uncertainty for the signal extraction varies from  1\% to 3\% (1\% to 5\%).  
The single muon tracking and trigger efficiencies, $\epsilon_{trk}$  and $\epsilon_{trg}$,  are estimated with the embedded \jpsi\ simulation.
The centrality dependence of these quantities is weak, since the decrease in most central collisions is about  1\% and 3.5\% for $\epsilon_{trk}$ and $\epsilon_{trg}$ respectively.
A 11\% systematic uncertainty on the tracking efficiency is estimated by comparing its determination based on real data and on a MC  approach. 
This estimation relies on a calculation of the tracking efficiency in each station using the  detector redundancy (two independent detection planes per station).   
The single track efficiency, defined as product of the station efficiencies, is calculated for tracks from real data and from simulation.  
The single track efficiencies are then injected in pure \jpsi\ simulations and the difference used as the \jpsi\ tracking systematic uncertainty. 
The  \pt\ and \y\ dependence of the former uncertainty leads to a bin to bin uncorrelated component ranging between 8\% and 14\%.
The systematic uncertainty on the \jpsi\  \aeps\ corrections related to the trigger efficiency is 2\%, mostly given by the uncertainty on the intrinsic efficiency of the trigger chambers. 
The systematic uncertainty related to the response function of  the trigger is always below 1\% except in the lowest \jpsi\  \pt\ interval where a value of 3\% was estimated. 
As a function of centrality, the systematic uncertainty of the  tracking or the  trigger efficiencies is 1\% in the most central collisions and becomes negligible for peripheral  collisions.   
The uncertainty on the matching efficiency between tracks reconstructed in the  tracking and trigger chambers amounts to 1\%. 
It is correlated as a function of the centrality and uncorrelated as a function of \pt\ and \y.

The \aeps\ calculation depends on the \jpsi\ \pt\ and \y\  distributions used as an input to the MC, and systematic effects originating from different parameterizations of these distributions must be taken into account. 
In the \ee\ analysis, the \jpsi\ (\pt, \y) parameterization is based on an interpolation of the RHIC, CDF and LHC data in \pp\ and p$\bar{\mathrm{p}}$ collisions \cite{Bossu:2011qe} corrected using nuclear shadowing calculations \cite{Eskola:1998df}.
The systematic uncertainty was evaluated by varying the slope of the \pt\ shape in a wide range such that the
average \pt\ changes between 1.5\gevc\ and 3.0\gevc\ . The \jpsi\ \pt\ spectra in A-A collisions 
measured by PHENIX \cite{Adare:2011yf} at mid-rapidity and ALICE at forward rapidity~\cite{Abelev:2012kr}
at all available centralities, together with their uncertainties, are well covered in the envelope determined 
by the considered variations. 
A centrality correlated systematic uncertainty of 5\% was obtained following this procedure.   
In the \mumu\ analysis, the MC \jpsi\ parameterizations are based on the  \pt\ and \y\   distributions measured  for different centrality classes. The  \pt\ -- \y\ correlation observed by LHCb in \pp\ collisions~\cite{Aaij:2011jh}  was also included in the systematic study.
A correlated variation in  \aeps\  of  3\%  was observed as a function of centrality. 
The \pt\ (\y) dependence of this systematic uncertainty brings a maximum contribution of 1\% (8\%) on each point.

Further sources of systematic uncertainties affecting the nuclear modification factor are the uncertainty on the limits of the centrality classes~\cite{Abelev:2013qoq}, on the  nuclear overlap function and on the \jpsi\ cross-section in pp collisions  at \sqrtSE{2.76}.  
An uncertainty on the  normalization factor $F_{\mathrm{norm}}$, accounting for run by run fluctuations on this quantity, is also added in the muon analysis.
All numerical values can be found in Table~\ref{tab:syst}.

When computing the \aeps\ factor, we assumed that \jpsi\ are produced unpolarized and no systematic uncertainty is assigned to a possible polarization. 
In \pp\ collisions, mid-rapidity ($\pt > 10 \gevc$) and forward-rapidity ($\pt > 2 \gevc$) measurements  have been done at \sqrtSE{7} and indeed  show that \jpsi\ polarization is compatible with zero~\cite{Chatrchyan:2013cla,Abelev:2011md,Aaij:2013nlm}.
In \PbPb\ collisions, \jpsi\ mesons produced from charm quarks in the medium are expected to be unpolarized.

\section{Results}
\label{sec:raaresults}

In the \ee\ decay channel, the inclusive \jpsi\ \Raa\ was studied as a function of the collision centrality (0\%--10\%,   10\%--40\% and 40\%--90\%)  for  $ \pt > 0$\gevc\ and  $ |\y| < 0.8$ .
In the \mumu\ decay channel, the event sample collected in the 2011 run  with the dedicated \mumuMB\ trigger allows for the study of the \Raa\ as a function of the centrality of the collisions in nine intervals.
Furthermore, a differential study of the \Raa\  as a function of transverse momentum or rapidity is also feasible.
Data are analyzed in seven intervals in the \pt\ range  $0< \pt <8$\gevc\ range and six intervals in the \y\ range $2.5< \y <4$.
The chosen binning matches the one adopted for the \sqrtSE{2.76}  pp results,  which are used as the reference for the evaluation of the nuclear modification factor. 

\begin{figure} [h!]
\begin{center}
\includegraphics[width=0.8\linewidth,keepaspectratio]{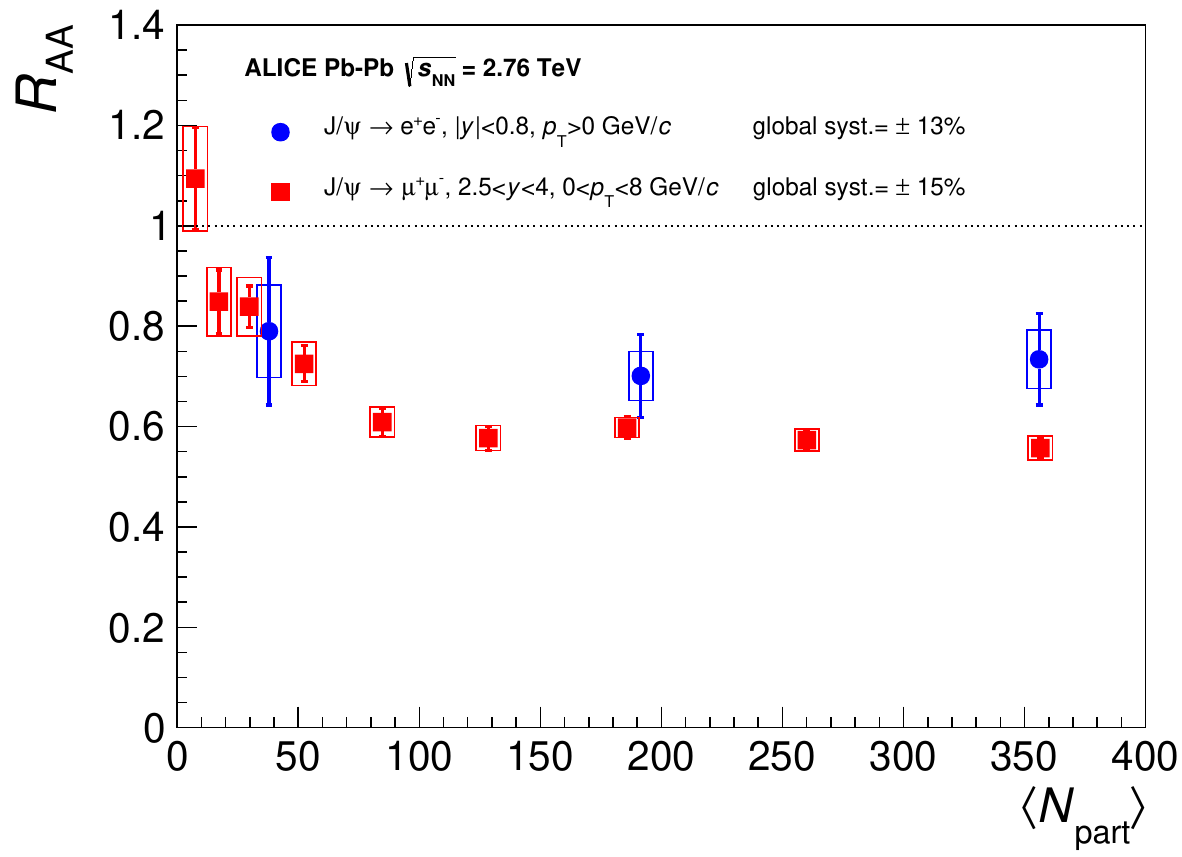}
\caption{(Color online) \label{fig:raaexp}
Centrality dependence of the nuclear modification factor, \Raa,  of inclusive \jpsi\ production in \PbPb\ collisions at  \sqrtSnnE{2.76}, measured at mid-rapidity and at forward-rapidity.
The point to point uncorrelated systematic uncertainties (type II) are represented as boxes around the data points,  while the statistical ones are shown as vertical bars. 
Global correlated systematic uncertainties (type I) are quoted directly in the legend.}
\end{center}
\end{figure}

Fig.~\ref{fig:raaexp} shows the inclusive \jpsi\ \Raa\ at mid- and forward-rapidity as a function of the number of  participant nucleons \Npart .
Statistical uncertainties are shown as vertical error bars, while the boxes represent the various uncorrelated  systematic uncertainties added in  quadrature. 
The systematic uncertainties correlated bin by bin (type II in Table~\ref{tab:syst})  are summed in quadrature and referred to as {\it global syst.} in the legend.
At forward-rapidity a clear suppression is observed,  independent of centrality for $\Npart > 70$.
Although with larger uncertainties, the mid-rapidity \Raa\  shows  a suppression of the \jpsi\  yield too.
The centrality integrated \Raa\ values are $\Raa^{0\%-90\%} = 0.72 \pm 0.06 (\mathrm{stat.}) \pm 0.10  (\mathrm{syst.})$ and
$\Raa^{0\%-90\%} = 0.58 \pm 0.01 (\mathrm{stat.}) \pm 0.09 (\mathrm{syst.})$ at mid- and forward-rapidity, respectively.
The systematic uncertainties on both \Raa\ values include the contribution arising from \Taa\ calculations.  
This amounts to 3.4\% of the computed \Taa\ value and is a  correlated systematic uncertainty common to the mid- and forward-rapidity measurements.
PHENIX mid- ($| \y | < 0.35$) and forward-rapidity ($1.2 < | \y | < 2.2$) results on inclusive \jpsi\ \Raa\ at \sqrtSnnE{0.2} exhibit a much stronger dependence on the collision centrality and a suppression of about a factor of three larger in the most central collisions~\cite{Adare:2011yf}. 

The measured inclusive \jpsi\ \Raa\ includes  contributions from prompt and non-prompt \jpsi ; the first one results from  direct \jpsi\  production  and feed-down  from \psip\ and \chic , the second one  arises  from beauty hadron decays.  
Non-prompt \jpsi\ are different with respect to the prompt ones, since their suppression or production  is insensitive  to color screening or regeneration mechanisms. 
Beauty hadron decay mostly occurs outside the fireball, and a measurement of the non-prompt \jpsi\ \Raa\ is  therefore  connected to  the beauty quark in-medium energy loss (see~\cite{Armesto:2005iq} and references therein).
At mid-rapidity, the contribution from beauty hadron feed-down to the inclusive \jpsi\ yield in \pp\ collisions at  \sqrtSE{7} is approximately 15\% \cite{Abelev:2012gx}. 
The prompt \jpsi\ \Raa\ can be evaluated according to $\Raa^{prompt} = (\Raa  - \Raa^{non-prompt}) / ( 1 -F_{\mathrm{B}} )$ where  $F_{\mathrm{B}}$ is the fraction of non-prompt \jpsi\  measured in pp collisions, and $\Raa^{non-prompt}$ is the nuclear modification factor of beauty hadrons in \PbPb\ collisions.
Thus,  the prompt \jpsi\ \Raa\ at mid-rapidity is expected to be about 7\% smaller than the inclusive measurement  if the beauty production scales with the number of binary collisions ($\Raa^{non-prompt} = 1$) and about 17\% larger if the beauty is fully suppressed ($\Raa^{non-prompt} = 0$). 
At forward-rapidity,  the non-prompt \jpsi\ fraction was measured  by the LHCb Collaboration to be about 11(7)\%  in \pp\ collisions  at \sqrtSE{7(2.76)} in the \pt\ range covered by this analysis~\cite{Aaij:2011jh,Aaij:2012asz}. 
Then, the difference between the \Raa\ of prompt \jpsi\ and the one for inclusive \jpsi\ is expected to be of about $-6$\%  and 7\%  in the two aforementioned extreme cases assumed  for beauty production.

\begin{figure}[htp!]
\begin{center}
\includegraphics[width=0.8\linewidth,keepaspectratio]{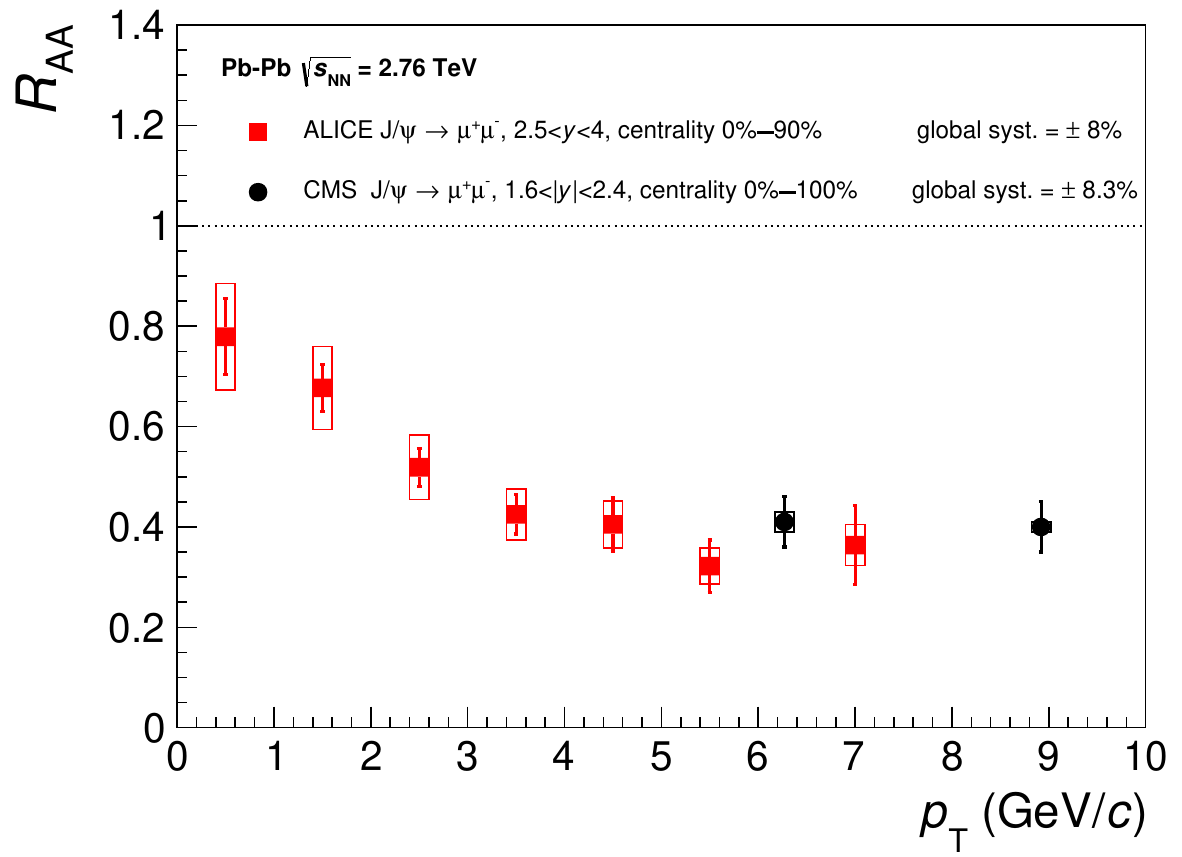}
\includegraphics[width=0.8\linewidth,keepaspectratio]{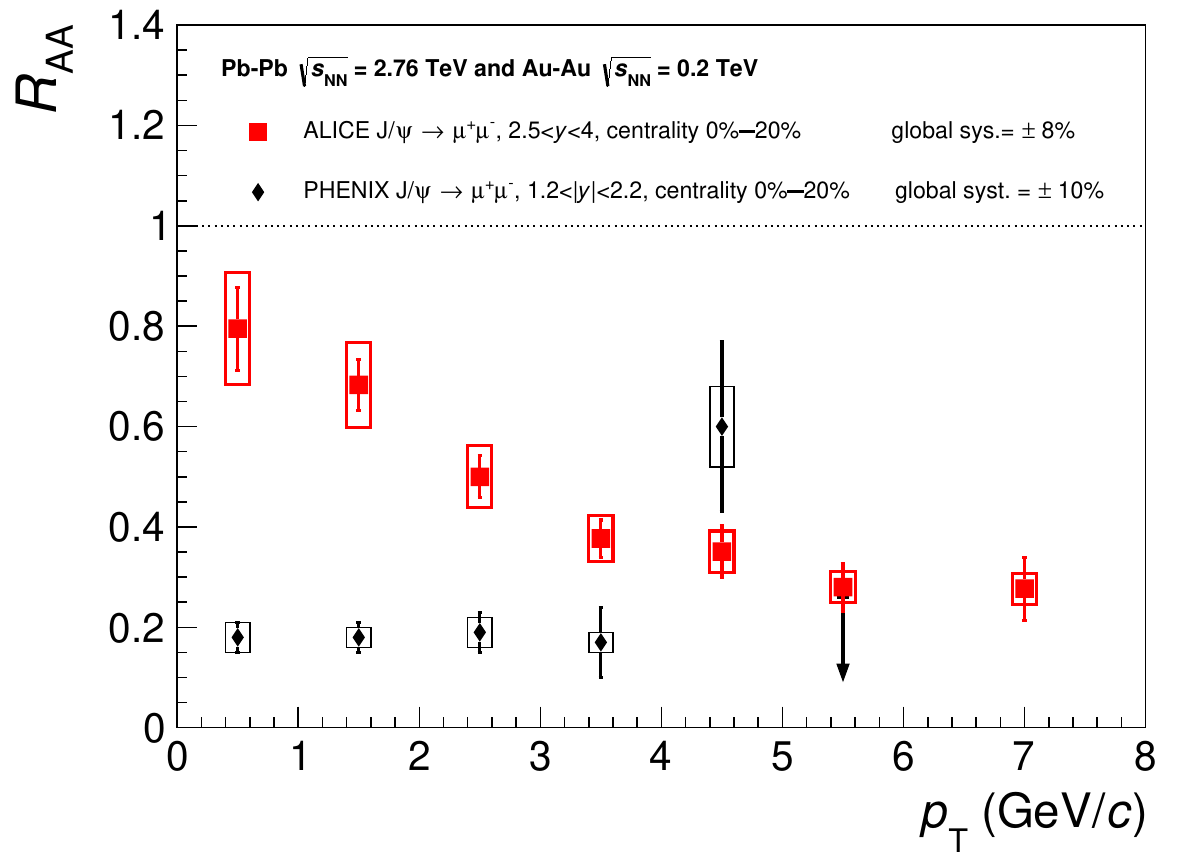}
\caption{(Color online) \label{fig:raaexp2} 
Top panel: transverse momentum dependence of the centrality integrated \jpsi\ \Raa\ measured by ALICE in \PbPb\ collisions at \sqrtSnnE{2.76} compared to CMS~\cite{Chatrchyan:2012np} results at the same \sqrtSnn. 
Bottom panel: transverse momentum dependence of the \jpsi\ \Raa\ measured by ALICE in the 0\%--20\% most central \PbPb\ collisions at \sqrtSnnE{2.76} compared to PHENIX~\cite{Adare:2011yf} results in the 0\%--20\% most central \AuAu\ collisions at \sqrtSnnE{0.2}.}
\end{center}
\end{figure}

In the top panel of Fig.~\ref{fig:raaexp2}, the  \jpsi\ \Raa\ at forward-rapidity is shown as a function of \pt\ for the 0\%--90\% centrality integrated  \PbPb\ collisions. 
It exhibits a decrease from 0.78 to 0.36, indicating that high \pt\ \jpsi\ are more suppressed than low \pt\ ones.    
Furthermore, at high \pt\, a direct comparison with CMS results~\cite{Chatrchyan:2012np} at the same \sqrtSnn\ is possible,  the main difference being that the CMS measurement covers a slightly more central rapidity range ($ 1.6  < |\y| < 2.4 $).
In the overlapping \pt\ range a similar suppression is found.
One should add here that the two CMS points are not independent and correspond to different intervals of the \jpsi\ \pt\  ($3 < \pt < 30\gevc$ and  $6.5 < \pt < 30\gevc$).
In the bottom panel of Fig.~\ref{fig:raaexp2}, the forward-rapidity \jpsi\ \Raa\ for the 0\%--20\% most central collisions is shown.
The observed \pt\ dependence of the \Raa\ for most central collisions is very close to the one in the 0\%--90\% centrality class. 
This is indeed expected since almost 70\% of the \jpsi\ yield  is contained in the 0\%--20\% centrality class.
Our data are compared to results obtained by PHENIX in 0\%--20\% most central \AuAu\ collisions at \sqrtSnnE{0.2}, in the rapidity region $1.2 < |\y| < 2.2$~\cite{Adare:2011yf}.
A striking difference between the \jpsi\ \Raa\ patterns can be observed. 
In particular, in the low \pt\ region the ALICE \Raa\ result is a factor of up to four higher compared to the PHENIX one.
This observation is in qualitative agreement with the calculations from~\cite{Zhao:2007hh,Zhou:2013} where the (re)combination dominance  in the \jpsi\ production leads to a decrease of the \meanptsq\ in \nucnuc\ collisions with respect to \pp\ collisions.
Although at the two energies the rapidity coverages are not the same and CNM effects might have a different size, our results point to the presence of a new contribution to the \jpsi\ yield at low \pt.

\begin{figure}[htp!]
\begin{center}
\includegraphics[width=0.8\linewidth,keepaspectratio]{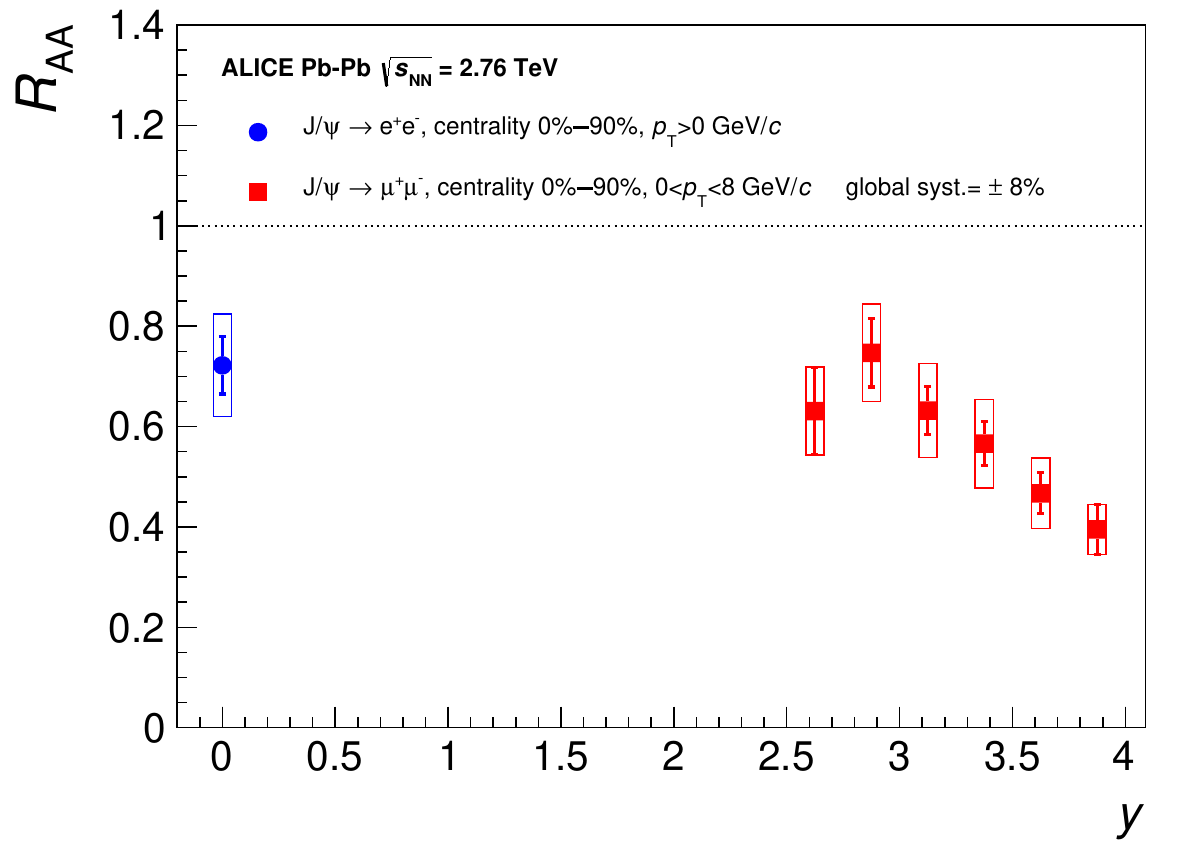}
\caption{(Color online) \label{fig:raaexp3} The rapidity dependence of the  \jpsi\ \Raa\ measured in  \PbPb\ collisions at \sqrtSnnE{2.76}. Both mid- and forward-rapidity measurements include a common correlated systematic uncertainty of 3.4\%  due to \Taa.
The mid-rapidity measurement covers the rapidity range $ | \y | < 0.8 $ and the forward-rapidity one is given in intervals of 0.25 unit of rapidity from $\y = 2.5 $ to $ \y = 4 $.}
\end{center}
\end{figure}

Finally, the  dependence of the \jpsi\ \Raa\ on rapidity is displayed in Fig.~\ref{fig:raaexp3} for  the 0\%--90\% centrality class.
At forward-rapidity, the  \jpsi\ \Raa\ decreases by about 40\% from $\y = 2.5$ to $\y = 4$. 
The result from the electron analysis is consistent with a constant or slightly increasing \Raa\ towards mid-rapidity.

\section{Conclusions}
\label{sec:conclusion}

The inclusive \jpsi\ nuclear modification factor has been measured by ALICE as a function of centrality,  \pt\ and \y\ in  \PbPb\ collisions at \sqrtSnnE{2.76}, down to zero \pt.
At  forward-rapidity, \Raa\ shows  a clear suppression  of the \jpsi\ yield, with no significant  dependence on centrality for \Npart\ larger than 70.   
At mid-rapidity, the \jpsi\ \Raa\ is compatible with a constant suppression as a function of centrality.
At forward-rapidity the \jpsi\ \Raa\ exhibits a strong \pt\ dependence and decreases by a factor of 2 from low \pt\ to high \pt.
This behavior strongly  differs from that observed by PHENIX  at \sqrtSnnE{0.2}.
This result suggests that a fraction of the \jpsi\ yield is produced via (re)combination of charm quarks.
In addition, the  indication of a non-zero \jpsi\ elliptic flow  in  \PbPb\ collisions at \sqrtSnnE{2.76} observed by ALICE~\cite{ALICE:2013xna} brings another hint in favor of (re)combination scenarios.
Precise knowledge of the cold nuclear effects is necessary for further understanding of the \jpsi\ behaviour.
The measurement of the \jpsi\ production in \pPb\ collisions at the LHC~\cite{Abelev:2013yxa,Aaij:2013zxa}  will allow one to sharpen the interpretation of these results.

\bibliographystyle{apsrev4-1}
\bibliography{./Raa_paper.bib}

\newenvironment{acknowledgement}{\relax}{\relax}
\begin{acknowledgement}
\section{Acknowledgments}
The ALICE collaboration would like to thank all its engineers and technicians for their invaluable contributions to the construction of the experiment and the CERN accelerator teams for the outstanding performance of the LHC complex.
\\
The ALICE collaboration gratefully acknowledges the resources and support provided by all Grid centres and the Worldwide LHC Computing Grid (WLCG) collaboration.
\\
The ALICE collaboration acknowledges the following funding agencies for their support in building and
running the ALICE detector:
 \\
State Committee of Science,  World Federation of Scientists (WFS)
and Swiss Fonds Kidagan, Armenia,
 \\
Conselho Nacional de Desenvolvimento Cient\'{\i}fico e Tecnol\'{o}gico (CNPq), Financiadora de Estudos e Projetos (FINEP),
Funda\c{c}\~{a}o de Amparo \`{a} Pesquisa do Estado de S\~{a}o Paulo (FAPESP);
 \\
National Natural Science Foundation of China (NSFC), the Chinese Ministry of Education (CMOE)
and the Ministry of Science and Technology of China (MSTC);
 \\
Ministry of Education and Youth of the Czech Republic;
 \\
Danish Natural Science Research Council, the Carlsberg Foundation and the Danish National Research Foundation;
 \\
The European Research Council under the European Community's Seventh Framework Programme;
 \\
Helsinki Institute of Physics and the Academy of Finland;
 \\
French CNRS-IN2P3, the `Region Pays de Loire', `Region Alsace', `Region Auvergne' and CEA, France;
 \\
German BMBF and the Helmholtz Association;
\\
General Secretariat for Research and Technology, Ministry of
Development, Greece;
\\
Hungarian OTKA and National Office for Research and Technology (NKTH);
 \\
Department of Atomic Energy and Department of Science and Technology of the Government of India;
 \\
Istituto Nazionale di Fisica Nucleare (INFN) and Centro Fermi -
Museo Storico della Fisica e Centro Studi e Ricerche "Enrico
Fermi", Italy;
 \\
MEXT Grant-in-Aid for Specially Promoted Research, Ja\-pan;
 \\
Joint Institute for Nuclear Research, Dubna;
 \\
National Research Foundation of Korea (NRF);
 \\
CONACYT, DGAPA, M\'{e}xico, ALFA-EC and the EPLANET Program
(European Particle Physics Latin American Network)
 \\
Stichting voor Fundamenteel Onderzoek der Materie (FOM) and the Nederlandse Organisatie voor Wetenschappelijk Onderzoek (NWO), Netherlands;
 \\
Research Council of Norway (NFR);
 \\
Polish Ministry of Science and Higher Education;
 \\
National Science Centre, Poland;
 \\
 Ministry of National Education/Institute for Atomic Physics and CNCS-UEFISCDI - Romania;
 \\
Ministry of Education and Science of Russian Federation, Russian
Academy of Sciences, Russian Federal Agency of Atomic Energy,
Russian Federal Agency for Science and Innovations and The Russian
Foundation for Basic Research;
 \\
Ministry of Education of Slovakia;
 \\
Department of Science and Technology, South Africa;
 \\
CIEMAT, EELA, Ministerio de Econom\'{i}a y Competitividad (MINECO) of Spain, Xunta de Galicia (Conseller\'{\i}a de Educaci\'{o}n),
CEA\-DEN, Cubaenerg\'{\i}a, Cuba, and IAEA (International Atomic Energy Agency);
 \\
Swedish Research Council (VR) and Knut $\&$ Alice Wallenberg
Foundation (KAW);
 \\
Ukraine Ministry of Education and Science;
 \\
United Kingdom Science and Technology Facilities Council (STFC);
 \\
The United States Department of Energy, the United States National
Science Foundation, the State of Texas, and the State of Ohio.
\end{acknowledgement}
\newpage
%
%
\appendix
\section{The ALICE Collaboration}
\label{app:collab}



\begingroup
\small
\begin{flushleft}
B.~Abelev\Irefn{org74}\And
J.~Adam\Irefn{org38}\And
D.~Adamov\'{a}\Irefn{org82}\And
M.M.~Aggarwal\Irefn{org86}\And
G.~Aglieri~Rinella\Irefn{org34}\And
M.~Agnello\Irefn{org92}\textsuperscript{,}\Irefn{org109}\And
A.G.~Agocs\Irefn{org132}\And
A.~Agostinelli\Irefn{org26}\And
N.~Agrawal\Irefn{org45}\And
Z.~Ahammed\Irefn{org128}
% ...
\renewcommand\labelenumi{\textsuperscript{\theenumi}~}

\section*{Affiliation notes}
\renewcommand\theenumi{\roman{enumi}}
\begin{Authlist}
\item \Adef{0}Deceased
\item \Adef{idp1136288}{Also at: St-Petersburg State Polytechnical University}
\item \Adef{idp3051840}{Also at: Department of Applied Physics, Aligarh Muslim University, Aligarh, India}
\item \Adef{idp3757824}{Also at: M.V. Lomonosov Moscow State University, D.V. Skobeltsyn Institute of Nuclear Physics, Moscow, Russia}
\item \Adef{idp4009200}{Also at: University of Belgrade, Faculty of Physics and "Vin\v{c}a" Institute of Nuclear Sciences, Belgrade, Serbia}
\item \Adef{idp4298624}{Permanent address: Konkuk University, Seoul, Korea}
\item \Adef{idp4830800}{Also at: Institute of Theoretical Physics, University of Wroclaw, Wroclaw, Poland}
\item \Adef{idp5738224}{Also at: the University of Kansas, Lawrence, KS, United States}
\end{Authlist}

\section*{Collaboration Institutes}
\renewcommand\theenumi{\arabic{enumi}~}
\begin{Authlist}

\item \Idef{org1}A.I. Alikhanyan National Science Laboratory (Yerevan Physics Institute) Foundation, Yerevan, Armenia
\item \Idef{org2}Benem\'{e}rita Universidad Aut\'{o}noma de Puebla, Puebla, Mexico
\item \Idef{org3}Bogolyubov Institute for Theoretical Physics, Kiev, Ukraine
\item \Idef{org4}Bose Institute, Department of Physics and Centre for Astroparticle Physics and Space Science (CAPSS), Kolkata, India
\item \Idef{org5}Budker Institute for Nuclear Physics, Novosibirsk, Russia
\item \Idef{org6}California Polytechnic State University, San Luis Obispo, CA, United States
\item \Idef{org7}Central China Normal University, Wuhan, China
\item \Idef{org8}Centre de Calcul de l'IN2P3, Villeurbanne, France
\item \Idef{org9}Centro de Aplicaciones Tecnol\'{o}gicas y Desarrollo Nuclear (CEADEN), Havana, Cuba
\item \Idef{org10}Centro de Investigaciones Energ\'{e}ticas Medioambientales y Tecnol\'{o}gicas (CIEMAT), Madrid, Spain
\item \Idef{org11}Centro de Investigaci\'{o}n y de Estudios Avanzados (CINVESTAV), Mexico City and M\'{e}rida, Mexico
\item \Idef{org12}Centro Fermi - Museo Storico della Fisica e Centro Studi e Ricerche ``Enrico Fermi'', Rome, Italy
\item \Idef{org13}Chicago State University, Chicago, USA
\item \Idef{org14}Commissariat \`{a} l'Energie Atomique, IRFU, Saclay, France
\item \Idef{org15}COMSATS Institute of Information Technology (CIIT), Islamabad, Pakistan
\item \Idef{org16}Departamento de F\'{\i}sica de Part\'{\i}culas and IGFAE, Universidad de Santiago de Compostela, Santiago de Compostela, Spain
\item \Idef{org17}Department of Physics and Technology, University of Bergen, Bergen, Norway
\item \Idef{org18}Department of Physics, Aligarh Muslim University, Aligarh, India
\item \Idef{org19}Department of Physics, Ohio State University, Columbus, OH, United States
\item \Idef{org20}Department of Physics, Sejong University, Seoul, South Korea
\item \Idef{org21}Department of Physics, University of Oslo, Oslo, Norway
\item \Idef{org22}Dipartimento di Fisica dell'Universit\`{a} 'La Sapienza' and Sezione INFN Rome
\item \Idef{org23}Dipartimento di Fisica dell'Universit\`{a} and Sezione INFN, Cagliari, Italy
\item \Idef{org24}Dipartimento di Fisica dell'Universit\`{a} and Sezione INFN, Trieste, Italy
\item \Idef{org25}Dipartimento di Fisica dell'Universit\`{a} and Sezione INFN, Turin, Italy
\item \Idef{org26}Dipartimento di Fisica e Astronomia dell'Universit\`{a} and Sezione INFN, Bologna, Italy
\item \Idef{org27}Dipartimento di Fisica e Astronomia dell'Universit\`{a} and Sezione INFN, Catania, Italy
\item \Idef{org28}Dipartimento di Fisica e Astronomia dell'Universit\`{a} and Sezione INFN, Padova, Italy
\item \Idef{org29}Dipartimento di Fisica `E.R.~Caianiello' dell'Universit\`{a} and Gruppo Collegato INFN, Salerno, Italy
\item \Idef{org30}Dipartimento di Scienze e Innovazione Tecnologica dell'Universit\`{a} del  Piemonte Orientale and Gruppo Collegato INFN, Alessandria, Italy
\item \Idef{org31}Dipartimento Interateneo di Fisica `M.~Merlin' and Sezione INFN, Bari, Italy
\item \Idef{org32}Division of Experimental High Energy Physics, University of Lund, Lund, Sweden
\item \Idef{org33}Eberhard Karls Universit\"{a}t T\"{u}bingen, T\"{u}bingen, Germany
\item \Idef{org34}European Organization for Nuclear Research (CERN), Geneva, Switzerland
\item \Idef{org35}Excellence Cluster Universe, Technische Universit\"{a}t M\"{u}nchen, Munich, Germany
\item \Idef{org36}Faculty of Engineering, Bergen University College, Bergen, Norway
\item \Idef{org37}Faculty of Mathematics, Physics and Informatics, Comenius University, Bratislava, Slovakia
\item \Idef{org38}Faculty of Nuclear Sciences and Physical Engineering, Czech Technical University in Prague, Prague, Czech Republic
\item \Idef{org39}Faculty of Science, P.J.~\v{S}af\'{a}rik University, Ko\v{s}ice, Slovakia
\item \Idef{org40}Frankfurt Institute for Advanced Studies, Johann Wolfgang Goethe-Universit\"{a}t Frankfurt, Frankfurt, Germany
\item \Idef{org41}Gangneung-Wonju National University, Gangneung, South Korea
\item \Idef{org42}Gauhati University, Department of Physics, Guwahati, India
\item \Idef{org43}Helsinki Institute of Physics (HIP), Helsinki, Finland
\item \Idef{org44}Hiroshima University, Hiroshima, Japan
\item \Idef{org45}Indian Institute of Technology Bombay (IIT), Mumbai, India
\item \Idef{org46}Indian Institute of Technology Indore, Indore (IITI), India
\item \Idef{org47}Inha University, College of Natural Sciences
\item \Idef{org48}Institut de Physique Nucleaire d'Orsay (IPNO), Universite Paris-Sud, CNRS-IN2P3, Orsay, France
\item \Idef{org49}Institut f\"{u}r Informatik, Johann Wolfgang Goethe-Universit\"{a}t Frankfurt, Frankfurt, Germany
\item \Idef{org50}Institut f\"{u}r Kernphysik, Johann Wolfgang Goethe-Universit\"{a}t Frankfurt, Frankfurt, Germany
\item \Idef{org51}Institut f\"{u}r Kernphysik, Technische Universit\"{a}t Darmstadt, Darmstadt, Germany
\item \Idef{org52}Institut f\"{u}r Kernphysik, Westf\"{a}lische Wilhelms-Universit\"{a}t M\"{u}nster, M\"{u}nster, Germany
\item \Idef{org53}Institut Pluridisciplinaire Hubert Curien (IPHC), Universit\'{e} de Strasbourg, CNRS-IN2P3, Strasbourg, France
\item \Idef{org54}Institute for High Energy Physics, Protvino, Russia
\item \Idef{org55}Institute for Nuclear Research, Academy of Sciences, Moscow, Russia
\item \Idef{org56}Institute for Subatomic Physics of Utrecht University, Utrecht, Netherlands
\item \Idef{org57}Institute for Theoretical and Experimental Physics, Moscow, Russia
\item \Idef{org58}Institute of Experimental Physics, Slovak Academy of Sciences, Ko\v{s}ice, Slovakia
\item \Idef{org59}Institute of Physics, Academy of Sciences of the Czech Republic, Prague, Czech Republic
\item \Idef{org60}Institute of Physics, Bhubaneswar, India
\item \Idef{org61}Institute of Space Science (ISS), Bucharest, Romania
\item \Idef{org62}Instituto de Ciencias Nucleares, Universidad Nacional Aut\'{o}noma de M\'{e}xico, Mexico City, Mexico
\item \Idef{org63}Instituto de F\'{\i}sica, Universidad Nacional Aut\'{o}noma de M\'{e}xico, Mexico City, Mexico
\item \Idef{org64}iThemba LABS, National Research Foundation, Somerset West, South Africa
\item \Idef{org65}Joint Institute for Nuclear Research (JINR), Dubna, Russia
\item \Idef{org66}Konkuk University, Seoul, South Korea
\item \Idef{org67}Korea Institute of Science and Technology Information, Daejeon, South Korea
\item \Idef{org68}KTO Karatay University, Konya, Turkey
\item \Idef{org69}Laboratoire de Physique Corpusculaire (LPC), Clermont Universit\'{e}, Universit\'{e} Blaise Pascal, CNRS--IN2P3, Clermont-Ferrand, France
\item \Idef{org70}Laboratoire de Physique Subatomique et de Cosmologie (LPSC), Universit\'{e} Joseph Fourier, CNRS-IN2P3, Institut Polytechnique de Grenoble, Grenoble, France
\item \Idef{org71}Laboratori Nazionali di Frascati, INFN, Frascati, Italy
\item \Idef{org72}Laboratori Nazionali di Legnaro, INFN, Legnaro, Italy
\item \Idef{org73}Lawrence Berkeley National Laboratory, Berkeley, CA, United States
\item \Idef{org74}Lawrence Livermore National Laboratory, Livermore, CA, United States
\item \Idef{org75}Moscow Engineering Physics Institute, Moscow, Russia
\item \Idef{org76}National Centre for Nuclear Studies, Warsaw, Poland
\item \Idef{org77}National Institute for Physics and Nuclear Engineering, Bucharest, Romania
\item \Idef{org78}National Institute of Science Education and Research, Bhubaneswar, India
\item \Idef{org79}Niels Bohr Institute, University of Copenhagen, Copenhagen, Denmark
\item \Idef{org80}Nikhef, National Institute for Subatomic Physics, Amsterdam, Netherlands
\item \Idef{org81}Nuclear Physics Group, STFC Daresbury Laboratory, Daresbury, United Kingdom
\item \Idef{org82}Nuclear Physics Institute, Academy of Sciences of the Czech Republic, \v{R}e\v{z} u Prahy, Czech Republic
\item \Idef{org83}Oak Ridge National Laboratory, Oak Ridge, TN, United States
\item \Idef{org84}Petersburg Nuclear Physics Institute, Gatchina, Russia
\item \Idef{org85}Physics Department, Creighton University, Omaha, NE, United States
\item \Idef{org86}Physics Department, Panjab University, Chandigarh, India
\item \Idef{org87}Physics Department, University of Athens, Athens, Greece
\item \Idef{org88}Physics Department, University of Cape Town, Cape Town, South Africa
\item \Idef{org89}Physics Department, University of Jammu, Jammu, India
\item \Idef{org90}Physics Department, University of Rajasthan, Jaipur, India
\item \Idef{org91}Physikalisches Institut, Ruprecht-Karls-Universit\"{a}t Heidelberg, Heidelberg, Germany
\item \Idef{org92}Politecnico di Torino, Turin, Italy
\item \Idef{org93}Purdue University, West Lafayette, IN, United States
\item \Idef{org94}Pusan National University, Pusan, South Korea
\item \Idef{org95}Research Division and ExtreMe Matter Institute EMMI, GSI Helmholtzzentrum f\"ur Schwerionenforschung, Darmstadt, Germany
\item \Idef{org96}Rudjer Bo\v{s}kovi\'{c} Institute, Zagreb, Croatia
\item \Idef{org97}Russian Federal Nuclear Center (VNIIEF), Sarov, Russia
\item \Idef{org98}Russian Research Centre Kurchatov Institute, Moscow, Russia
\item \Idef{org99}Saha Institute of Nuclear Physics, Kolkata, India
\item \Idef{org100}School of Physics and Astronomy, University of Birmingham, Birmingham, United Kingdom
\item \Idef{org101}Secci\'{o}n F\'{\i}sica, Departamento de Ciencias, Pontificia Universidad Cat\'{o}lica del Per\'{u}, Lima, Peru
\item \Idef{org102}Sezione INFN, Bari, Italy
\item \Idef{org103}Sezione INFN, Bologna, Italy
\item \Idef{org104}Sezione INFN, Cagliari, Italy
\item \Idef{org105}Sezione INFN, Catania, Italy
\item \Idef{org106}Sezione INFN, Padova, Italy
\item \Idef{org107}Sezione INFN, Rome, Italy
\item \Idef{org108}Sezione INFN, Trieste, Italy
\item \Idef{org109}Sezione INFN, Turin, Italy
\item \Idef{org110}SUBATECH, Ecole des Mines de Nantes, Universit\'{e} de Nantes, CNRS-IN2P3, Nantes, France
\item \Idef{org111}Suranaree University of Technology, Nakhon Ratchasima, Thailand
\item \Idef{org112}Technical University of Split FESB, Split, Croatia
\item \Idef{org113}Technische Universit\"{a}t M\"{u}nchen, Munich, Germany
\item \Idef{org114}The Henryk Niewodniczanski Institute of Nuclear Physics, Polish Academy of Sciences, Cracow, Poland
\item \Idef{org115}The University of Texas at Austin, Physics Department, Austin, TX, USA
\item \Idef{org116}Universidad Aut\'{o}noma de Sinaloa, Culiac\'{a}n, Mexico
\item \Idef{org117}Universidade de S\~{a}o Paulo (USP), S\~{a}o Paulo, Brazil
\item \Idef{org118}Universidade Estadual de Campinas (UNICAMP), Campinas, Brazil
\item \Idef{org119}University of Houston, Houston, TX, United States
\item \Idef{org120}University of Jyv\"{a}skyl\"{a}, Jyv\"{a}skyl\"{a}, Finland
\item \Idef{org121}University of Liverpool, Liverpool, United Kingdom
\item \Idef{org122}University of Tennessee, Knoxville, TN, United States
\item \Idef{org123}University of Tokyo, Tokyo, Japan
\item \Idef{org124}University of Tsukuba, Tsukuba, Japan
\item \Idef{org125}University of Zagreb, Zagreb, Croatia
\item \Idef{org126}Universit\'{e} de Lyon, Universit\'{e} Lyon 1, CNRS/IN2P3, IPN-Lyon, Villeurbanne, France
\item \Idef{org127}V.~Fock Institute for Physics, St. Petersburg State University, St. Petersburg, Russia
\item \Idef{org128}Variable Energy Cyclotron Centre, Kolkata, India
\item \Idef{org129}Vestfold University College, Tonsberg, Norway
\item \Idef{org130}Warsaw University of Technology, Warsaw, Poland
\item \Idef{org131}Wayne State University, Detroit, MI, United States
\item \Idef{org132}Wigner Research Centre for Physics, Hungarian Academy of Sciences, Budapest, Hungary
\item \Idef{org133}Yale University, New Haven, CT, United States
\item \Idef{org134}Yonsei University, Seoul, South Korea
\item \Idef{org135}Zentrum f\"{u}r Technologietransfer und Telekommunikation (ZTT), Fachhochschule Worms, Worms, Germany
\end{Authlist}
\endgroup

\end{document}